\documentclass[12pt,preprint]{aastex}
\usepackage{psfig}

\def\hnot{\ifmmode H_0 \else H$_0$ \fi}

\def\msun{\ifmmode {\rm M_\odot} \else M$_\odot$\fi}
\def\lsun{\ifmmode {\rm L_\odot} \else L$_\odot$\fi}

\def\deg{\ifmmode ^{\circ}
         \else $^{\circ}$\fi}
\def\pdeg{\ifmmode 
           $\setbox0=\hbox{$^{\circ}$}\rlap{\hskip.11\wd0 .}$^{\circ}
     \else \setbox0=\hbox{$^{\circ}$}\rlap{\hskip.11\wd0 .}$^{\circ}$\fi}

\def\arcsec{\ifmmode '' \else $''$\fi}
\def\arcsecpt{\ifmmode ''\!\!. \else $''\!\!.$\fi}

\def\msunyr{\ifmmode {\rm M_\odot~yr^{-1}}\else${\rm M_\odot~yr^{-1}}$\fi}
\def\lam{\ifmmode {\lambda} \else {$\lambda$} \fi}

\def\mdoto{\ifmmode {\dot{M}_0} \else  $\dot{M}_0$ \fi}
\def\teff{\ifmmode {T_{eff}} \else $T_{eff}$ \fi}
\def\ilam{\ifmmode {I_\lambda} \else  $I_\lambda$ \fi}
\def\inu{\ifmmode {I_\nu} \else  $I_\nu$ \fi}
\def\fnu{\ifmmode {F_\nu} \else  $F_\nu$ \fi}

\def\yr{\ifmmode {\rm yr} \else  yr \fi}
\def\cm{\ifmmode {\rm cm} \else  cm \fi}
\def\cmmitwo{\ifmmode \rm cm^{-2} \else $\rm cm^{-2}$\fi}
\def\cmmithree{\ifmmode \rm cm^{-3} \else $\rm cm^{-3}$\fi}
\def\cmps{\ifmmode \rm cm~s^{-1}\else $\rm cm~s^{-1}$\fi}
\def\cmpsps{\ifmmode \rm cm~s^{-2}\else $\rm cm~s^{-2}$\fi}
\def\kmps{\ifmmode \rm km~s^{-1}\else $\rm km~s^{-1}$\fi}
\def\kmpspmpc{\ifmmode \rm km~s^{-1}~Mpc^{-1} \else
    $\rm km~s^{-1}~Mpc^{-1}$\fi}
\def\ergps{\ifmmode \rm erg~s^{-1} \else $\rm erg~s^{-1}$ \fi}
\def\ergpspcm{\ifmmode \rm erg~s^{-1}~cm^{-2} \else $\rm erg~s^{-1}~cm^{-2}$ \fi}
\def\ergpspcmphz{\ifmmode \rm erg~s^{-1}~cm^{-2}~Hz^{-1} \else $\rm
erg~s^{-1}~cm^{-2}~Hz^{-1}$ \fi}
\def\ergpspcmpa{\ifmmode \rm erg~s^{-1}~cm^{-2}~\AA^{-1} \else $\rm
erg~s^{-1}~cm^{-2}~\AA^{-1}$ \fi}
\def\ergpsphz{\ifmmode \rm erg s^{-1} Hz^{-1} \else 
   $\rm erg s^{-1} Hz^{-1}$ \fi} 

\def\mbh{\ifmmode M_{BH} \else $M_{BH}$ \fi}
\def\mbhsig{$M_{BH} - \sigma$}
\def\mbhsigstar{\ifmmode M_{BH} - \sigma_* \else $M_{BH} - \sigma_*$ \fi}

\def\mbhsigthree{$M_{BH} - \sigma_{[O~III]}$}
\def\sigstar{\ifmmode \sigma_* \else $\sigma_*$ \fi}
\def\sigthree{\ifmmode \sigma_{[O~III]} \else $\sigma_{[O~III]}$ \fi}
\def\wthree{\ifmmode {\rm FWHM([O~III])} \else $FWHM([O~III])$ \fi}
\def\mthree{\ifmmode M_{[O~III]} \else $M_{[O~III]}$ \fi}
\def\hbeta{\ifmmode {\rm H}\beta \else H$\beta$ \fi}
\def\oiii{\rm [O~III]}
\def\log{\rm{log}}

\newcommand{\feii}{Fe\,{\sc ii}}
\newcommand{\hb}{H{\sc $\beta$}}

\def\etal{~et al.}

\accepted{}
\journalid{}{}
\articleid{}{}

\slugcomment{Submitted to The Astrophysical Journal}

\shortauthors{G. Shields}
\shorttitle{Black Holes}
\received{2002 July 3}
\begin{document}

\title{The Black Hole--Bulge Relationship in QSOs}

\author{
Gregory A. Shields\altaffilmark{1},
Karl Gebhardt\altaffilmark{1},
Sarah Salviander\altaffilmark{1},
Beverley J. Wills\altaffilmark{1},
Bingrong Xie\altaffilmark{2},
Michael S. Brotherton\altaffilmark{3},
Juntao Yuan\altaffilmark{1},
and Matthias Dietrich\altaffilmark{4}
}

\altaffiltext{1}{Department of Astronomy, University of Texas, Austin, TX
78712; shields@astro.as.utexas.edu, gebhardt@astro.as.utexas.edu, 
triples@astro.as.utexas.edu, bev@astro.as.utexas.edu, juntao@astro.as.utexas.edu}

\altaffiltext{2}{
 Department of Physics and Astronomy,
   Rutgers, The State University of New Jersey,
   P.O. Box 849, Piscataway, NJ 08855-0849;
   xiebr@physics.rutgers.edu}

\altaffiltext{3}{National Optical Astronomy Observatories,
  950 N. Cherry Ave., Tucson, AZ 85726; mbrother@ohmah.tuc.noao.edu}

\altaffiltext{4}{Department of Astronomy, University of Florida, 211 Bryant
Space Science Center, Gainesville, FL 32611-2055; dietrich@astro.ufl.edu}

\begin{abstract}

We use QSO emission-line widths to examine the \mbhsigstar\
relationship as a function of redshift and to extend the relationship to larger
masses. 
Supermassive black holes in galactic nuclei are closely related to the bulge of
the host galaxy.  The
mass of the black hole, $\mbh$, increases with the bulge luminosity and with
the velocity dispersion of the bulge stars,
$\sigstar$. An important clue to the origin of this correlation
would be an observational
determination of the evolution, if any, in the \mbhsigstar\ relationship as a function
of cosmic time.  The high luminosity of
QSOs affords the potential for studies at large redshifts. 
We derive black hole masses from the continuum luminosity and
the width of the broad H$\beta$ line and  $\sigstar$ from the
width of the narrow [O~III] lines.   We
find that radio quiet QSOs conform to the established \mbhsigstar\
relationship up to values $\mbh \approx\ 10^{10}~\msun$, with no discernible
change in the relationship out to redshifts of $z \approx 3$.  These results are
consistent with the idea that the growth of supermassive black holes and massive bulges
occurred simultaneously.

\end{abstract}


\section{INTRODUCTION}

The evolution of supermassive black holes and that of their
host galaxies appear to be closely
coupled.  Every bulge system studied with high spatial resolution shows a central black
hole, presumably a relic of AGN activity (Kormendy~\& Richstone
1995; Richstone~\etal\ 1999; Kormendy \& Gebhardt 2001).  The black
hole mass is roughly proportional to the luminosity of the bulge of the host
galaxy, albeit with scatter of $\sim 0.5$~dex in \mbh\ (Magorrian~\etal\ 1998).  
Gebhardt~\etal\ (2000a) and Ferrarese~\& Merritt (2000) found a
tighter correlation involving \mbh\ and $\sigstar$, where $\sigstar$ is the stellar
line-of-sight velocity dispersion at radii outside the gravitational influence of the
black hole.  Tremaine et al. (2002) give this relationship as 
\begin{equation}
\mbh = (10^{8.13}~\msun)(\sigstar/200~\kmps)^{4.02}.
\end{equation}
with an intrinsic scatter
$\leq 30\%$.
(The measure of scatter here is 1~$\sigma$ for \mbh at fixed bulge
luminosity or $\sigstar$, respectively.)
Theoretical interpretations of
this correlation (e.g., Silk~\& Rees 1998; Adams~\etal\ 2001; Burkert~\& Silk
2001; Ostriker 2001; Balberg~\& Shapiro 2002; Haehnelt~\& Kauffmann 2000)
variously would have the black hole form before, during, or after bulge
formation.   Given this uncertainty, measurements of \mbh\ in galaxies with
large look-back times would be valuable.  Most promising would be a
measurement of the \mbhsigstar\ relationship at high redshift, because the relatively
small scatter of this relationship in the local universe should make it possible to
discern even modest changes over time.

Unfortunately, measurements of \mbh\ and $\sigstar$ at high redshift are
challenging because spatially resolved kinematical studies are feasible only for nearby
systems.  However, recent work indicates that useful measures of \mbh\ in AGN are
possible (see Peterson 1997; Wandel, Peterson, \& Malkan 1999).  McLure~\& Dunlop
(2001) and Wandel (2001) summarize the ``reverberation'' and ``photoionization''
methods to measure \mbh\ in Seyfert galaxies. The results correlate with bulge
luminosity in a way consistent with nearby galaxies, confirming the calibration of \mbh
for AGN.  However, the host galaxy brightness is difficult to measure in the presence
of a bright active nucleus, and a direct measurement of
$\sigstar$ for quasar host galaxies using stellar absorption-lines is
likewise difficult.  Alternatively, the narrow emission-line widths can serve as a
surrogate for $\sigstar$.  
Nelson~\& Whittle (1996) find
agreement between narrow  emission-line widths in AGNs and widths measured from
stellar absorption-line kinematics.  On this basis, Nelson (2000) proposes
to use the width of the [OIII]
$\lambda\lambda 5007, 4959$ lines of AGN as a surrogate for $\sigstar$.  Thus, 
one takes $\sigstar \approx \sigthree$, where $\sigthree \equiv \wthree/2.35$
and FWHM is the full width at half maximum.  The divisor 2.35 relates $\sigma$ and
FWHM for a Gaussian profile.  Taking black hole masses from reverberation mapping,
Nelson finds no systematic offset in the
\mbhsigthree\ relationship between low redshift AGN measured this way and measurements
of galaxies based on observed stellar velocity dispersions.

In this paper, we use emission-line widths to study the \mbhsigthree\
relationship for a sample
of quasars at redshifts up to 3.3.  Our goal is to assess the
evolutionary history of black hole growth compared with that of the host
galaxy.  In Section 2, we 
describe the derivation of \mbh\ and the use of the [O~III] line width as a
surrogate for $\sigstar$. Then we describe our adopted data set,
which relies on published data covering a range of redshift together
with our unpublished observations of low redshift QSOs. In Section 3, we present the
results, and we examine the trend of \mbh\ with
$\sigthree$ and the redshift dependence of this
relationship.  In Section 4, we discuss these results and the needed
improvements.  Throughout our discussion, we use a cosmology with
$\hnot = 70~\kmpspmpc, \Omega_{\rm M} = 0.3, \Omega_{\Lambda} = 0.7$.
All values of luminosity used in this paper are corrected to our adopted
cosmological parameters.

\section{METHOD}

\subsection{Calculation of Black Hole Masses}

The derivation of black hole masses from AGN broad line widths has been
discussed by a number of authors, recently including Laor (1998), Wandel \etal\ (1999),
Kaspi \etal\ (2000), Vestergaard (2002), and  McLure \& Dunlop (2001).  The method
relies on assuming that the line widths, at least for some lines, are dominated
by orbital motion of the emitting gas in the gravitation potential of the black
hole.  This is supported by the
generally symmetrical time variability of the wings, and the decrease in line
width with increasing radius for different lines in a given object (see Wandel
\etal\ 1999, and references therein).  The black hole mass is then given by 
$\mbh = v^2R/G$, where $v$ and $R$ are an appropriate velocity
and radius for the BLR.  Deriving $v$ and $R$ from the observations is not
entirely straightforward, as the emitting gas spans a range of radii and
velocities.  Pragmatically, the radius is derived from ``reverberation mapping''
or ``echo mapping'' studies that monitor the variation of the continuum and
emission lines.  The characteristic time lag between continuum variations and the
response of a given line gives a measure of the radius of the region emitting
that line (see review by Peterson 1993).  Given this determination of $R$, one
needs a measure of line width which, combined with this radius, gives the
correct black hole mass.  This choice can be parameterized as
$v = f\times \rm FWHM $ for the line.  Some authors use $f = \sqrt{3}/2$,
appropriate for isotropic velocities.  However, McLure \& Dunlop (2001)
argue that allowance for a flattened geometry of the BLR
is preferable.   Most work has employed the Balmer
lines, in particular \hbeta.   Black
hole masses determined in this way for AGN with measured
$\sigstar$ show overall agreement with the
\mbhsigstar\ relationship (Gebhardt \etal\ 2000b; Ferrarese \etal\ 2001; McLure \&
Dunlop 2001).

The work of Wandel \etal\ (1999) and Kaspi \etal\ (2000) generated a set of
BLR radii based on variability of the Balmer lines for 17 Seyfert 1 (Sy 1)
galaxies and 17 PG QSOs.  This work supports earlier indications that the BLR
radius increases with luminosity, approximately as $R \propto L^{0.5}$.  
Photoionization physics suggests $R \propto L^{0.5}$, and this is also 
consistent with the idea that the BLR radius may be
limited by the minimum radius at which dust grains survive (Netzer \& Laor 1993). 
This relationship is enormously useful, for it opens the door to measuring black hole
masses in large numbers of AGN from measurements of line width and luminosity at a
single epoch.  By determining the BLR radius in this way, one avoids the need for long
series of observations of variability, which are in any case impractical for luminous,
high redshift QSOs with long variability timescales.  The use of such ``photoionization
masses'', calibrated in terms of echo results, has been discussed by several authors,
including Wandel \etal\ (1999) and Vestergaard (2002).  There is some
controversy over the slope of the radius-luminosity relationship.  Kaspi \etal\
(2000) find $R \propto L^{0.7}$, based on echo radii.  However,  
McLure \& Jarvis (2002; see also Maoz 2002), find  $R \propto L^{0.61}$.  On this
basis, they fit the echo masses with
$
M_{BH} = (10^{7.63}~\msun)v_{3000}^2 L_{44}^{0.61}, 
$
where $v_{3000} \equiv FWHM(\hbeta)/3000~\kmps$ and
$L_{44} \equiv \lambda L_{\lambda}(5100~{\rm \AA})/(10^{44}~\ergps)$, using the same
cosmology that we have adopted.  The difference in slope results from several factors,
including a new echo mass for NGC 4051, different continuum luminosities, and the
cosmological model.  Here we take for our primary calibration the physically motivated
slope $R \propto L^{0.5}$.  We adopt the $L^{0.5}$ fit shown in Figure 6 of Kaspi \etal,
\begin{equation}
M_{BH} = (10^{7.69}~\msun)v_{3000}^2 L_{44}^{0.5}, 
\end{equation}
where we have adjusted the coefficient to our cosmology.
(This expression agrees with our own fitting of the Kaspi et al data and depends little on
NGC 4051 because the slope is fixed.)

We use the \hbeta\ line width in this fashion to determine \mbh, adopting the
calibration of equation (2) for most of our discussion but illustrating the key results
also for $M_{BH} \propto L^{0.61}$.  McLure
\& Jarvis and Vestergaard (2002) discuss the use of other lines.
The use of the $\lambda5100$ continuum follows Wandel \etal\ (1999) and Kaspi
\etal\ (2000).  McLure
\& Jarvis (2002) and Laor \etal\ (1997) argue that the $\lambda3000$~\AA\ continuum may
work better, but the differences are not great.  The $\lambda 5100$
continuum has the practical advantage that it can be measured in the same
spectra as the \hbeta and \oiii\ lines, and it is less affected by dust extinction.

\subsection{[O~III] Lines in AGN}

Nelson \& Whittle (1996) have compared [O~III] line widths and stellar
velocity dispersions in AGN, finding generally good agreement.  For the quantity log
$\sigthree/\sigstar$, they find a mean $0.00\pm 0.01$ and a dispersion $\sigma
= 0.20$, supporting the idea that the motions of the NLR gas are largely determined by
the gravitational potential of the host galaxy.  This is reinforced by the analysis
by Nelson (2000), who shows essentially that \sigthree\ and \mbh\ obey equation (1) for
AGN with echo values of $\mbh$.  These results support the use of
$\sigthree$ as a surrogate for $\sigstar$.  A caution, however, 
is that [O~III] profiles often
have substantial asymmetry and non-Gaussian profiles, possibly resulting from
outflow combined with extinction of the far side of the NLR (e.g., 
Wilson \& Heckman 1985; Nelson \& Whittle 1995).

Radio loud AGN tend to
have stronger [O~III] emission than radio quiet objects, as reflected in
``Eigenvector 1'' of Boroson \& Green (1992).  Radio jets may contribute to
the motions of the NLR gas (Nelson \& Whittle 1996).  For this reason, we emphasize
radio quiet objects in this paper, and we discuss radio loud objects separately.

\subsection{New Observations}

We include here results from an unpublished set of spectra of QSOs obtained
at McDonald Observatory with the Large Cassegrain Spectrograph (LCS) on the 2.7-meter
telescope.  These objects correspond to the X-ray sample studied by Laor et al. (1997),
which is in turn taken from the Bright Quasar Survey  based on
the Palomar Green (PG) survey (Boroson \& Green 1992; Schmidt \& Green 1983).  We
take values of
FWHM($\hbeta$) from Boroson \& Green.  The instrumental
resolution was 150 to 180~\kmps\ FWHM, depending on redshift.  
Emission from Fe II blends was removed from the spectra with the aid of the
Boroson \& Green (1992) template.
Table 1 gives the \oiii\
widths for 2 RL and 14 RQ objects in the redshift range 0.09 to 0.33 derived from our
spectra.  We use here the results of directly 
measuring the half-maximum point of the
observed line profile rather than by fitting any kind of curve. 
The width of $\lambda4959$ was noted for corroboration.
(See below for a discussion of methods of measuring the [O~III] line width.)
Our own measurements of FWHM for
\hbeta\ ranged from 0.95 to 1.35 times the BG92 values, with a mean ratio of 1.09.  This
may give some indication of systematic uncertainties arising from Fe II subtraction and
fitting procedure, although real temporal variations 
may affect individual objects.  For our
PG sample, continuum luminosities were taken from Laor \etal\ (1997).  Table 2
lists the redshift, adopted line widths, continuum luminosity, and black hole mass
derived from equation (2).

\subsection{Data from the literature}

We have drawn observations of H$\beta$ and \oiii\ from several published
sources.  These are listed below, roughly in order of decreasing redshift.  The results
are given in Table 2. The quoted line widths are intrinsic values after subtraction of the
instrumental width.

(1)~Dietrich et al. (2002, ``D02'') present infrared spectra for six QSOs with
$z \approx 3.4$, placing \hbeta\ and \oiii\ in the infrared $K$-band wavelength
region.  Of these, Q0256-0000 and Q0302-0019 have 
\oiii\ lines of adequate strength to measure the line width.  We have measured
the FWHM of \oiii\ and
\hbeta\ from the original data, using a direct measurement of the FWHM.  We subtract in
quadrature the instrumental width of 400~\kmps. We find $\wthree = 838\pm29~\kmps$ for
Q0256-0000 and
$743\pm19~\kmps$ for Q0302-0019, where the quoted errors reflect only the noise
in the data.  (For a single Gaussian fit to the [O~III] profiles, we
find 886 and 846~\kmps, respectively. These are 0.02, 0.05~dex larger than the direct
measurements.  The sense of the difference is typical, but the magnitude is not
significant for our purposes. The appropriateness of Gaussian profiles is discussed
below.)  We 
corrected for the narrow component of $\hbeta$ by assuming a typical ratio of $\lambda 5007$
to narrow \hbeta\ of 10 to 1, as discussed in Brotherton (1996b, ``B96b'') and
references therein.  The quoted noise errors and the likely error resulting from
subtraction of the narrow \hbeta\ component are small compared to
scatter among objects described below.  Continuum luminosities are discussed
below.

(2)~McIntosh et al. (1999, ``M99'') give
$H$-band spectra of 32 luminous QSOs at $2.0 \leq z \leq 2.5$, covering \hbeta\
and \oiii.  The instrumental resolution is about 500~\kmps. The signal-to-noise ratio
(S/N) of the spectra is poor in some cases.  Therefore, we present in Figures 1 and 2 two
different data selections.  For the ``full" sample, we include all objects for which M99
give FWHM for both \hbeta\ and [O~III].  For the ``select'' sample, we include those QSOs
for which the greatest of the positive and negative errors in FWHM for both \hbeta\ and 
[O~III], as a fraction of FWHM, is 0.33 or less.  This results in a list of 4 radio
loud (RL) and 4 radio quiet (RQ) objects.  The ``full'' sample offers a larger number of
objects for averaging purposes, and it avoids the question of bias in selecting the
better measurements.  However, it contains some rather uncertain measurements and may
exaggerate the true dispersion of the data.  The ``select'' sample is consistent with a
reasonable judgment of the most reliable measurements, based on the S/N of the spectra.  (The
spectra are given in M99.)  M99 quote a FWHM for the ``total
H$\beta$'' profile and for the ``broad H$\beta$'' component.  The ``broad $\hbeta$'' width
exceeds the ``total'' width by more than can be accounted for by the removal of a narrow
\hbeta\ component with a typical intensity ratio to \oiii\ (see above). 
We have carried out our own fits to the M99 data for some representative objects,
including  a broad \hbeta\ component plus a narrow
\hbeta\ component with the same width as the [O~III] lines.  We find that the
width of the broad \hbeta\ component is fairly close to M99's ``total \hbeta''
width and much narrower than their quoted ``broad $\hbeta$'' width.  Accordingly,
we use FWHM for $\lambda5007$ and H$\beta_{total}$ from Table 4 of M99 and a RL/RQ type from
their Table 2.  See Vestergaard (2002) for a discussion of the appropriate measure of \hbeta\
width for determinations of black hole mass.  Average errors in \wthree\ are $\pm
0.12, 0.16$~dex for the RL, RQ objects in the ``full'' sample and 0.07, 0.10~dex for the
``select'' sample, respectively.  Errors are in most cases smaller for \hbeta\ than [O~III]
width.  Error bars are omitted from our figures for clarity, and the reader should bear these
errors in mind.

(3)~Brotherton (1996a, ``B96a'') gives results of infrared spectroscopy of 18
RL and 14 RQ QSOs ranging in redshift from 0.7 to 2.5.  Of these, 11 RQ and
7 RL objects were observed with the Cryogenic Spectrometer (CRSP) on the
2.1 meter telescope at Kitt Peak National Observatory (KPNO).  The remaining
objects were observed with other setups giving inadequate spectral resolution,
or are included in B96b.  All our
data sources give line widths corrected for instrumental resolution except B96a.  For
B96a, we subtract in quadrature the instrumental resolution of 470, 600, and
460~\kmps\ for I, J, and H band observations, respectively, as estimated by Brotherton (2002). 
Our ``full" data set includes all 11 CRSP radio quiet objects.  For the
``select" sample, we use all objects for which the fractional error in \wthree  is
$\leq 0.10$ (Brotherton
2002).  This is based on the error in the corrected line widths resulting from
the formal error in the original Gaussian fits by B96a.  These errors are not directly
comparable with those of M99, which involve template fitting and a Monte Carlo error
analysis.  However, the chosen cutoff agrees with a reasonable judgment of the most
reliable measurements, based on the S/N of the spectra.  Among the radio quiet
objects, the selected ones comprise four of the six CRSP objects for which B96a gives a quality
flag of ``A".  The select RL sample consists of 4 of the 5 CRSP ``A'' objects.  (The fifth
is PKS 0424-131, which is also in the M99 select sample.  We include the B96a
measurement in Table 2 for comparison.)  The [O~III] widths
tabulated by B96a involve single Gaussian fits.  From the spectra, we have estimated
\wthree\ by direct measurement and find typical agreement within about 10\% with the adopted
Gaussian fits for the RL and RQ select objects.  B96a also made no correction for the NLR
contribution to the \hbeta\ profile.  None of the spectra shows a prominent narrow line
component of
\hbeta.  For a typical ratio of
$\lambda 5007$ to narrow \hbeta\ of 10 to 1, the correction would on average increase
FWHM(H$\beta$) by 11\% for the RL select objects and 7\% for the RQ select objects.  These
differences are not significant for our purposes.   In addition to FWHM for
\hbeta\ and \oiii\, we took from Chapter 6 of B96a the apparent $V$ magnitudes and
RL/RQ classifications of the subject QSOs.  We note that 1120+01 is a
lens candidate, although it may be a true binary QSO (Kochanek 2002). 
Michalitsianos et al. (1997) suggest that the amplification factor could be as large as
100.  Correction for lensing would lower
\mbh\ as $L^{0.5}$ from the value in Table 2 and Figure 1.  Perhaps the \mbhsigthree\
relation has potential as an indicator of lensing.

(4)~B96b studied profiles of \hbeta\ and
\oiii\ in 60 radio loud quasars with $0.05 \leq z \leq 0.93$.  
The instrumental resolution was 150 to 300~\kmps.  We used the 
FWHM for \hbeta\ and
\oiii\ from his Tables 3 and 4.  Brotherton gives a
quality flag A or B for 37 objects.  We judged 6 of these to be unsuitable by visual
inspection of the spectra, typically because the broad \hbeta\ line was weak or had
an unusual profile.  (The spectra are given in in B96b.)  We also eliminated
those objects for which we could not obtain a flux density at 5100~\AA\ rest wavelength from
available spectra, as described below.  This left a sample of 23 QSOs that are included in
Table 2.

(5)~Grupe et
al. (1999, ``G99'') discuss optical emission-line properties of 76 bright soft X-ray
selected AGN, based on spectra with a resolution of $\sim 5$~\AA\ FWHM ($\sim300$~\kmps). 
Redshifts and line widths are taken from Table 1 of G99. 
The [O~III] line width was measured using a single Gaussian fit.
We selected the 33 objects for which the fractional uncertainty in 
FWHM was better than 0.10 both for
\hbeta and [O~III].  Four of these were eliminated as unsuitable on the basis of
inspection of the spectra as reproduced in Grupe (1996), and one for radio confusion.    
Two of the remaining objects were found from NED
\footnote{The NASA/IPAC Extragalactic Database (NED) is operated by the Jet
Propulsion Laboratory, California Institute of Technology, under contract
with the National Aeronautics and Space Administration.
} 
to be radio loud, leaving a sample of 26 radio quiet objects.

\subsection{Continuum Luminosity}

The luminosities for our samples come from heterogeneous sources.  An assumed power law
$F_\nu \propto \nu^{-0.5}$ was used when necessary to scale the measurements to
5100~\AA\ rest wavelength.  Magnitudes were converted to flux densities
$F_\nu$ as prescribed by Allen (1973).  When absolute magnitudes
or specific luminosities were given, these were adjusted to our adopted cosmology.

For D02, the continuum flux density at
5100~\AA\ rest wavelength was measured from the published spectra.  For M99 we used
$L_\nu(\rm V)$ from their Table 2, and for B96a we used the apparent V magnitude given
in his Chapter 6.
For B96b we used flux densities taken from spectrophotometry described by
Netzer et al. (1995) and  Wills et al. (1995) and unpublished spectrophotometry 
(Wills and Brotherton 2002) when
available and otherwise took them from Figure 1 of B96b.  Objects were omitted when
neither of these was available, or when the result differed by more than 0.5~dex from
the flux density at 5100~\AA\ estimated  from the absolute visual magnitude
$M_{abs}$ quoted by B96b (taken from V\'eron-Cetty \& V\'eron 1989).
For G99, we measured the optical continuum flux density from the
published spectra.  

We have not attempted a correction for a contribution to the continuum from
the host galaxy.  This
is unlikely to have a large effect on our conclusions, because \mbh\ scales
only as $L_{\nu}^{1/2}$.  We have estimated the host galaxy luminosity
from the value of $\sigthree$ for our individual AGN, using the 
Faber-Jackson relation (Forbes \& Ponman 1999; Bernardi et al. 2001), 
$M_V = -20.57 - 2.5\,[4\, \rm log(\sigthree/220~\kmps)]$.  For the adopted RL and RQ
objects in the D02, M99, B96, B96b, and PG samples, the average value of $L_{\rm
bulge}/L_{\rm observed}$ is only 2 or 3 percent, and 7 percent for G99. Thus, we
have assumed that the measured flux is dominated by the AGN continuum for the adopted
spectra of all objects.

\subsection{Radio Loudness}

Objects in the sample were divided into radio-quiet and radio-loud using the
radio-to-optical ratio, $R_{ro} = F_{\nu}(\mathrm{5 \
GHz})/F_{\nu}(4400\mathrm{\AA})$ as defined by Kellermann et al. (1989),
where $R_{ro} > 10$ designates a radio-loud QSO.  For our PG sample, we
used values for $R_{ro}$ from Boroson \& Green (1992); for B96a we used values from
that reference; for D02 we used values from Hooper et al. (1995); and
for M99 we used values from that paper.

For G99 we obtained flux densities at 5 GHz, 4.85
GHz, or 1.4 GHz from photometric points in NED; otherwise, we measured the
flux density from NRAO VLA Sky Survey (NVSS) radio flux density maps at 1.4
GHz (Condon et al. 1998). 
We assumed unresolved single, or very occasionally,
double unresolved components.  The angular resolution of NVSS of 45\arcsec\
(Condon et al. 1998) includes the extended structure for nearly all QSOs.
 When no detection was listed or
evident on the maps, we assumed that the object was radio quiet. For radio flux
densities at frequencies other than 5 GHz we extrapolate to
$F_{\nu}(\mathrm{5 \ GHz})$ assuming
$F_{\nu} \propto
\nu^{-0.7}$. Subtleties in the definition of $R_{ro}$ and continuum slope
have little effect for our purposes, because of wide range in $R_{ro}$
values between RQ and RL objects.  For purposes of computing $R_{ro}$,
optical continuum flux densities were found as described above, or taken from
NED, and corrected to rest wavelength $\lambda 4400$ using $L_{\nu} \propto \nu^{-0.5}$.

\section{RESULTS}

\subsection{The $\mathbf{M_{BH} - \sigma_{[O~III]}}$ relationship}

Figures 1 and 2 show results for the RQ objects, based on equation (2).   Figure 1
includes results for the full samples described above for the M99 and B96a data sets,
and Figure 2 shows the select samples.  (The data for the other sources is the same.)  
On average, the QSO results show good agreement with equation (1), shown as the straight
line in Figures 1 and 2.   This agreement supports the use of photoionization masses for
\mbh\ and the use of the [O~III] line width as a surrogate for $\sigstar$ to masses over
$10^{9}~\msun$.  However, we emphasize that the close alignment of the massive, high
redshift objects in the upper right part of Figure 2 is better than expected from the
errors (see discussion above) and must be to some degree fortuitous.  The scatter in the
PG sample may give a better indication of
the scatter resulting from measurement errors together with true variations among
objects, as discussed below.  We assign significance only to the average agreement of
the high redshift points as a group with the \mbhsigstar\ trend.

We also show in Figures 1 and 2
the objects listed in Table 1 of Nelson (2000).  The Nelson points
agree in the mean with
our results but show less scatter around the Tremaine et al.
(2002) line.  The echo masses used by Nelson are presumably more accurate 
than the photoionization
masses used here.  Also, Nelson restricted his study to data with spectral
resolution R $>$ 1500.  (Included in the Nelson points in Figures 1 and 2 are
two objects in common with our PG sample, namely 0953+414 and 1411+442.)

If the high redshift and Nelson (2000) points are removed from Figure 1, 
the scatter in the remaining points is too large to define accurately
the slope of the \mbhsigthree\ trend.  However, our philosophy here is that
the \mbhsigstar\ relationship is well established, and we are interested
in examining the agreement of the AGN data with that relationship.  The
lower redshift points agree in the mean with equation 1, and this supports
our working hypothesis that photoionization masses and \sigthree\ can
be used as a substitute for direct measurements of \mbh\
and \sigstar\ in AGN.  This is the basis for our discussion of the
redshift dependence below.

Nelson \& Whittle
(1996) found that AGN with powerful, linear radio sources sometimes have \wthree\ larger
than expected for the value of $\sigstar$. 
Figure 3 shows the radio loud objects from our data set, using the select sample for B96a
and M99.  Objects from most of the data sets scatter around equation (1).  However,
the measurements of B96b generally stand above and to the left
of the trend.   One issue may be the procedure for measuring the
FWHM of [O~III].  Some workers fit a Gaussian and quote the FWHM of this fit. 
B96b instead gives a direct measurement of the width at the half-maximum level of the
observed line profile, which is feasible only when the data have good spectral
resolution and signal-to-noise ratio. For four objects in common between the
single Gaussian fit of B96a and the direct measurement of B96b,
the B96b width on average is narrower by
$\sim0.2$~dex.  A single Gaussian fit may be affected by broad wings or
asymmetry of the line profile
and by a tendency to smooth the line peak.  This may explain some
of the displacement of the B96b points in Figure 2.  On the other hand, we also
made a direct measurement of \wthree\ for our PG sample, and in Figures 1 and 2 our
results do not share the displacement of the B96b sample.  Consistent with this, we
find that Gaussian fits to [O~III] for our PG sample are only 0.04~dex wider on average
than the direct measurements used here.  As a further illustration, we measured
\wthree\ using a Gaussian fit for many of the M99 objects, using the original data.  The
Gaussian widths are always wider than the M99 values, which should approximate a direct
measurement because of the multiple template fit used by M99.  For 3 RQ objects in our select
sample for M99, the Gaussians are wider by an average of 0.08~dex.  However, a few objects,
typically with weak or strongly asymmetrical [O~III], show differences in FWHM
approaching a factor 2.  Accurate results using [O~III] as a surrogate for
$\sigstar$ will require careful attention to the shape of the line profile.

\subsection{Redshift Dependence}

The  \mbhsigthree\ relationship shown in Figures 1 and 2 provides a basis for assessing
the evolution of the relationship with redshift.  The B96a, D02,
and M99 objects fall in the redshift range 1 to 3.3.  These
objects do not systematically depart from the overall \mbhsigthree\ relationship.  In
order to display the trend as a function of redshift, we use a fictitious
``[O~III] mass''  \mthree defined as the mass given by equation (1) with \sigthree
in place of \sigstar.  The quantity
$\Delta\, \log~M \equiv \log~M_{BH} - \log~\mthree$ is then a measure of the vertical
deviation of a given object from the Tremaine \etal\ (2002) mean trend in Figures 1 - 3.  
In Figures 4 and 5, we plot 
this quantity as a function of redshift for our full and select RQ samples, respectively.  
In the mean, the high
redshift objects show little deviation from the \mbhsigstar\ trend.  Because the most
massive objects are the high redshift objects,
we cannot
independently assess the slope of the \mbhsigthree\ relationship and the redshift dependence.
Our point is that {\em the data are consistent with a \mbhsigthree\ relationship that extends
to high masses with a slope close to that of Tremaine \etal\ (2002) and
that does not evolve strongly with time.}

A quantitative measure of the adherence of the high redshift objects to
the low redshift trend is difficult, given the small number of high redshift
objects in the select sample and the various systematic errors.  We may estimate the
intrinsic scatter of the high redshift samples on the basis of the PG sample, which
has the higher luminosity of
the low redshift samples.  The RQ objects in our PG sample show a mean $\Delta\, \log~M$
of +0.10 and a dispersion 0.5 ($1~\sigma$).    The mean of
$\Delta\, \log~M$ is -0.11
 for the 9 high redshift
RQ objects in the select sample.  The standard deviation of the mean for these 9
objects, if drawn from a population with similar dispersion to
the PG sample, would be about 0.2~dex. For the full RQ sample, there are 23 objects
with a mean $\Delta\, \log~M$ of -0.15, which suggests that there is no significant bias in
our select sample.  These results indicate that, at times corresponding to  redshifts $z
\approx 2$ to 3, the average
\mbh\ at a given \sigthree\ was within a factor 2 or 3 of the present day value.

One systematic uncertainty is the slope of the radius--luminosity relationship for
the BLR.  The high redshift RQ select objects are on average 1.8~dex
more luminous than the PG sample.  Use of the McLure and Jarvis 
(2002) relationship, with $R \propto L^{0.61}$, elevates the high redshift
points by $\sim0.2$~dex in Figures 1 - 5 while having little effect on the
low redshift samples.  This brings the high redshift and PG samples into even closer
agreement, with a mean
$\Delta\, \log~M$ of +0.15 for the RQ PG sample and +0.14 for the RQ high redshift select
sample. Figure 6 shows
$\Delta\, \log~M$ versus redshift for the McLure and Jarvis calibration (given before 
equation~2 above). Another uncertainty is the slope of the
\mbhsigstar\ relationship, discussed by Tremaine \etal\ (2002).  The high redshift
objects in Figure 1 have $\sigstar \approx 500~\kmps$.  If the slope were
$\mbh \propto \sigstar^{3.5}$ rather than
$\sigstar^{4.0}$, for example, the predicted mass for the high redshift objects would
be lowered by about 0.2~dex.  This would raise these objects by this
amount in Figures~4 - 6.

\section{DISCUSSION}

There are two main points that come from our analysis. First, Figures~1 and 2
show that we can place AGN on the \mbhsigstar\ correlation using the
[OIII] emission-line width as a surrogate for $\sigstar$. This bypasses
the need to obtain high signal-to-noise spectra in order to obtain the
host galaxy absorption line kinematics.  Nelson (2000) was the first to
point this out.  We confirm it with our independent data set including luminous QSOs and using
photoionization masses rather than reverberation masses.

Second, given that one can measure [OIII] line profiles at fairly
modest signal-to-noise levels in AGN, one can now study the
relationship between black holes and their host galaxies at high redshift.
Although a larger sample of high redshift objects is needed, Figures~4 - 6 suggests
that the \mbhsigthree\ relationship is not a strong function of redshift.   For the
adopted cosmology, the age of the universe was 2.0 Gyr at
$z = 3.3$ and 3.3 Gyr at $z = 2.0$.  The
objects in this redshift range have black hole masses up to $\sim 10^{10}~\msun$ 
and implied host galaxy masses up to $\sim 10^{13}~\msun$, if
$M_{\rm bulge} \approx 10^{2.8} \mbh$ (Kormendy \& Gebhardt 2001).   Our results suggest
that black holes typically grow contemporaneously with their host galaxy bulges, or else that
both were well formed by $z \approx 3$.

Recent results on black hole demographics suggest that high mass black holes acquire most of
their mass from luminous accretion during episodes of QSO activity (Yu \& Tremaine 2002,
and references therein).  Yu \& Tremaine find that half the black hole mass is accreted
before a redshift $z \approx 1.8$, and only 10\% before $z = 3$.  Thus, our results suggest
that the black hole-galaxy bulge relationship is roughly obeyed, at least by very massive
holes, at a time when much of the growth of present-day black holes lay
in the future.

Previous workers have noted a tendency for [O~III] widths to increase
with luminosity for AGN (e.g., B96b; M99; V\'eron-Cetty, V\'eron, \& Gon\c calves 2001).  In
the present context, this is a natural consequence of the tendency to have larger \mbh\ in
more luminous objects, together with the \mbhsig\ relationship.  B96b suggested that the
mass of the host galaxy might be involved in the increase of [O~III] width with
luminosity.

The conclusion that galaxy growth is contemporaneous with black hole growth
is consistent with chemical abundances in QSOs.  Heavy element
abundances in luminous QSOs are solar or several times solar
(Hamann \etal\ 2002).  The highest abundances are typically found in the
centers of large galaxies (e.g., Garnett \etal\ 1997).

We have crudely estimated bolometric luminosities for the AGN in our sample by
taking $L_{bol} = 9\nu L_{\nu}(\rm 5100~\AA)$ (Kaspi~\etal ~ 2000).  On average
the various data sets have ${\rm log} L_{bol}/L_{\rm Ed} \approx -0.4$, where
the Eddington limit $L_{\rm Ed}$ is calculated from $M_{\rm BH}(H\beta)$.
This ratio shows little systematic difference between the samples at higher and lower
luminosity or redshift.  Some individual objects exceed the Eddington limit on this
basis, but only by amounts that may be consistent with uncertainties in
\mbh\ and $L_{\rm bol}$.

The use of \sigthree\ provides a new way to estimate black hole masses in AGN.
Once the systematics of the [O~III] profile are understood in this context,
black hole masses from \sigthree\ may be as reliable as ``photoionization masses''
based on \hbeta\ width and continuum luminosity.  The availability of
two independent estimates of \mbh\ will allow workers to compare each for consistency
with other approaches, such as accretion disk fitting of the continuum energy
distribution (e.g., Mathur \etal\ 2001).   Both measures 
of \mbh\ may likewise be examined for
systematic correlations with various properties of AGN, such as the
correlation of  $L/L_{\rm Ed}$ with ``eigenvector 1'' found by
Boroson (2002).

The \mbhsigstar\ relationship offers, in principle, a new standard
candle for cosmology.  The value of \mbh\ derived from the broad \hbeta\ line width
varies as $L^a$, with $a \approx 0.5$ to 0.6 (see discussion preceding equation 2).
If one assumes that either \sigstar\ or \sigthree\ correctly predicts \mbh\ through
equation (1), then equation (2) can be solved for $L$ in terms of \sigstar\ and
FWHM(H$\beta$).  This gives
\begin{equation}
L_{44} = 10^{0.88} \sigma_{200}^{8.02} v_{3000}^{-4}.
\end{equation}
However, the use of equation (3) for cosmological measures will be challenging, 
given the large exponents of $\sigma_{200}$
and $v_{3000}$ and the uncertainties in the calibration 
of the \mbhsigthree\ and $M_{\rm BH}-\L$ relationships.

Future work on this topic should include a systematic examination of the
best way of characterizing the width of the [O~III] lines for use in 
predicting $\sigstar$ and calibration of the chosen measure.  Use of
shorter wavelength narrow lines such as [O II] would allow study of higher
redshift objects at a given observed wavelength.  An obvious need is for
more and better observations of the \hbeta\ and [O~III] region in high redshift
QSOs.  In order to minimize the uncertainty associated with the
scaling of BLR radius with luminosity, observations of high and low redshift
QSOs should be made as nearly as possible at similar intrinsic luminosity.
High resolution imaging studies may provide direct measures of the host galaxy
luminosity as a check on $\sigthree$.

\acknowledgments

G.A.S. gratefully acknowledges the hospitality of Lick Observatory during
summer 2001 and 2002. This material is based in part upon work supported by the Texas
Advanced Research Program under Grant No. 003658-0177-2001.
B.J.W. acknowledges financial support through NASA Long Term
Space Astrophysics grant NAG5-3431.
M.D. acknowledges support through NASA Long Term Space Astrophysics grant NAG5-3234 and
NSF research grant AST99-84040.  We are grateful to G. Blumenthal, T. Boroson, A. Gon\c
calves, D. Grupe, E. Hooper, J. Kormendy,  A. Laor, J. Miller, and D. Osterbrock for
helpful discussions.

\clearpage
\begin{deluxetable}{lccccc}
\tablewidth{0pt}
\tablecaption{[OIII] Widths of PG Quasars}
\tablehead{
\colhead{Name}       & 
\colhead{Date}       & 
\colhead{Integration}& 
\colhead{EW(FeII)}   & 
\colhead{\feii/\hb}  &
\colhead{FWHM([OIII]) } \\
\colhead{PG}         & 
\colhead{UT 1996}    & 
\colhead{seconds}    & 
\colhead{\AA}        & 
\colhead{}           &
\colhead{km s$^{-1}$}}
\startdata
0947$+$396  & Feb 14  & 2700 & 25 & 0.22 & 366$\pm$13 		\\
0953$+$414  & Feb 15  & 2100 & 39 & 0.30 & 571\phs 19 		\\
1001$+$054  & Feb 15  & 2700 & 73 & 0.74 & 678\phs \phn7 	\\
1048$+$342  & Feb 14  & 2700 & 73 & 0.57 & 286\phs \phn9 	\\
1114$+$445  & Feb 15  & 2400 & 20 & 0.19 & 538\phs \phn8 	\\
1115$+$407  & Feb 15  & 2700 & 33 & 0.42 & 224\phs 17 		\\
1115$+$407  & Feb 16  & 2700 & $\cdots$ & $\cdots$ & 267\phs 24 \\
1116$+$215  & Feb 15  & 2400 & 81 & 0.43 & 902\phs \phn7 	\\
1116$+$215  & Feb 16  & 2700 & $\cdots$ & $\cdots$ & 901\phs 18	\\
1202$+$281  & Feb 14  & 2100 & 20 & 0.12 & 412\phs \phn8 	\\
1216$+$069  & Feb 17  & 2400 &  8 & 0.04 & 343\phs 16 		\\
1226$+$023  & Feb 17  &  600 &  6 & 0.05 & 754\phs 20 		\\
1309$+$355  & Feb 14  & 2400 & 14 & 0.22 & 641\phs 12 		\\
1322$+$659  & Feb 14  & 2700 & 45 & 0.66 & 249\phs 15 		\\
1352$+$183  & Feb 14  & 2400 & 61 & 0.51 & 572\phs \phn8 	\\
1411$+$442  & Feb 15  & 2400 & 52 & 0.49 & 411\phs 10 		\\
1427$+$480  & Feb 16  & 2400 & 50 & 0.35 & 476\phs 28 		\\
1440$+$356  & Feb 15  & 1800 & 76 & 1.15 & 464\phs \phn8 	\\
1626$+$554  & Feb 14  & 2201 & 46 & 0.31 & 694\phs 23 		\\

\enddata

\tablecomments{Equivalent width, EW(FeII), is measured in the rest
frame using the Boroson \& Green (1992) \feii\ template. The full
width at half the maximum of the line profile, FWHM, is determined by
direct measurement from the \feii-corrected \oiii\ profile, not from
an analytical fit.  Values have been corrected by subtracting the
instrumental FWHM in quadrature.  Uncertainties are estimated rms,
including uncertainties from noise and in correcting for instrumental
resolution, but not including uncertainties in \feii\ subtraction.
The Fe II measurements refer to the region between 4434\AA\ and 4686\AA\
in the original I Zw 1  template constructed by Boroson \& Green (1992),
that is, they refer to a template with I Zw 1's line profile.
}

\end{deluxetable}

\begin{deluxetable}{llcccccc}
\tabletypesize{\small}
\tablewidth{0pt}
\tablenum{2}
\tablecaption{\mbh\ and \sigthree for AGN}
\tablehead{
\colhead{Name}          & 
\colhead{Alternate}     & 
\colhead{RL/RQ}         & 
\colhead{z}             &
\colhead{FWHM(H$\beta$)} & 
\colhead{log$(\nu L_\nu)$}    & 
\colhead{log(M$_{\rm BH}$)} & 
\colhead{log($\sigma_{\rm O III}$)} \\
\colhead{}          & 
\colhead{Name}      & 
\colhead{}          & 
\colhead{}          &
\colhead{km/s}      & 
\colhead{erg/s}     & 
\colhead{M$_\odot$} & 
\colhead{km/s}      }
\startdata
{\it D02}	  &  		  &  		  &  		  &  		  &  		  &  		  &  		   \\ 
0256-0000	  &  		  &  	L	  &  	3.377	  &  	4455	  &  	46.85	  &  	9.46	  &  	2.55	   \\ 
0302-0019	  &  		  &  	Q	  &  	3.286	  &  	3777	  &  	46.89	  &  	9.33	  &  	2.50	   \\ 
\\															
{\it M99}	  &  		  &  		  &  		  &  		  &  		  &  		  &  		   \\ 
0153+744	  &  		  &  	L	  &  	2.341	  &  	5650	  &  	47.16	  &  	9.82	  &  	2.70	   \\ 
0421+019	  &  		  &  	L	  &  	2.056	  &  	4660	  &  	46.63	  &  	9.39	  &  	2.77	   \\ 
0424-131	  &  		  &  	L	  &  	2.168	  &  	4380	  &  	46.50	  &  	9.27	  &  	2.70	   \\ 
1331+170	  &  		  &  	L	  &  	2.097	  &  	7480	  &  	46.98	  &  	9.97	  &  	2.88	   \\ 
0043+008	  &  		  &  	Q	  &  	2.146	  &  	4330	  &  	46.69	  &  	9.35	  &  	2.59	   \\ 
0109+022	  &  		  &  	Q	  &  	2.351	  &  	7020	  &  	46.68	  &  	9.77	  &  	2.77	   \\ 
1104-181	  &  		  &  	Q	  &  	2.318	  &  	3950	  &  	47.07	  &  	9.46	  &  	2.78	   \\ 
2212-179	  &  		  &  	Q	  &  	2.228	  &  	6150	  &  	46.76	  &  	9.69	  &  	2.71	   \\ 
\\															
{\it B96a}	  &  		  &  		  &  		  &  		  &  		  &  		  &  		   \\ 
0024+22	  &  	NAB	  &  	L	  &  	1.109	  &  	5883	  &  	46.34	  &  	9.45	  &  	2.70	   \\ 
0424-13	  &  	PKS	  &  	L	  &  	2.167	  &  	4818	  &  	46.60	  &  	9.40	  &  	2.78	   \\ 
0454+03	  &  	PKS	  &  	L	  &  	1.345	  &  	4896	  &  	46.54	  &  	9.39	  &  	2.46	   \\ 
0952+179	  &  	PKS	  &  	L	  &  	1.476	  &  	4574	  &  	46.35	  &  	9.23	  &  	2.52	   \\ 
1718+48	  &  	PG	  &  	L	  &  	1.082	  &  	4613	  &  	47.06	  &  	9.59	  &  	2.83	   \\ 
0117+213	  &  	PG	  &  	Q	  &  	1.499	  &  	6493	  &  	46.84	  &  	9.78	  &  	2.77	   \\ 
0920+580	  &  	SBS	  &  	Q	  &  	1.378	  &  	4928	  &  	46.18	  &  	9.21	  &  	2.47	   \\ 
1120+01	  &  	UM 425	  &  	Q	  &  	1.470	  &  	6821	  &  	46.64	  &  	9.72	  &  	2.75	   \\ 
1634+706	  &  	PG	  &  	Q	  &  	1.335	  &  	9601	  &  	47.21	  &  	10.30	  &  	2.94	   \\ 
\\															
{\it B96b}	  &  		  &  		  &  		  &  		  &  		  &  		  &  		   \\ 
0159-117	  &  	3C 057	  &  	L	  &  	0.670	  &  	4500	  &  	45.76	  &  	8.92	  &  	2.42	   \\ 
0414-060	  &  	3C 110	  &  	L	  &  	0.773	  &  	8200	  &  	45.91	  &  	9.52	  &  	2.39	   \\ 
0736+017	  &  	0	  &  	L	  &  	0.191	  &  	3400	  &  	44.74	  &  	8.17	  &  	2.36	   \\ 
0738+313	  &  	0	  &  	L	  &  	0.630	  &  	4800	  &  	45.60	  &  	8.90	  &  	2.19	   \\ 
0742+318	  &  	0	  &  	L	  &  	0.462	  &  	9940	  &  	45.74	  &  	9.60	  &  	2.29	   \\ 
0838+133	  &  	0	  &  	L	  &  	0.684	  &  	3000	  &  	45.40	  &  	8.39	  &  	2.19	   \\ 
0903+169	  &  	3C 215	  &  	L	  &  	0.411	  &  	4440	  &  	44.63	  &  	8.35	  &  	2.34	   \\ 
0923+392	  &  	0	  &  	L	  &  	0.698	  &  	7200	  &  	45.84	  &  	9.37	  &  	2.34	   \\ 
1004+130	  &  	PG	  &  	L	  &  	0.240	  &  	6300	  &  	45.22	  &  	8.94	  &  	2.30	   \\ 
1007+417	  &  	0	  &  	L	  &  	0.613	  &  	3560	  &  	45.61	  &  	8.65	  &  	2.34	   \\ 
1011+232	  &  	0	  &  	L	  &  	0.565	  &  	2700	  &  	45.23	  &  	8.21	  &  	2.17	   \\ 
1100+772	  &  	PG	  &  	L	  &  	0.311	  &  	6160	  &  	45.40	  &  	9.01	  &  	2.41	   \\ 
1103-006	  &  	PG	  &  	L	  &  	0.426	  &  	6560	  &  	45.26	  &  	9.00	  &  	2.52	   \\ 
1137+660	  &  	3C 263	  &  	L	  &  	0.652	  &  	6060	  &  	46.03	  &  	9.32	  &  	2.38	   \\ 
1150+497	  &  	0	  &  	L	  &  	0.334	  &  	4810	  &  	44.85	  &  	8.52	  &  	2.22	   \\ 
1217+023	  &  	0	  &  	L	  &  	0.240	  &  	4300	  &  	45.12	  &  	8.56	  &  	2.21	   \\ 
1250+568	  &  	SBS	  &  	L	  &  	0.320	  &  	4560	  &  	44.53	  &  	8.32	  &  	2.33	   \\ 
1305+069	  &  	3C 281	  &  	L	  &  	0.599	  &  	6440	  &  	45.28	  &  	9.00	  &  	2.46	   \\ 
1340+290	  &  	0	  &  	L	  &  	0.905	  &  	13000	  &  	45.83	  &  	9.88	  &  	2.28	   \\ 
1354+195	  &  	0	  &  	L	  &  	0.720	  &  	4400	  &  	45.95	  &  	9.00	  &  	2.30	   \\ 
1458+718	  &  	0	  &  	L	  &  	0.905	  &  	3000	  &  	45.64	  &  	8.51	  &  	2.46	   \\ 
1545+210	  &  	0	  &  	L	  &  	0.264	  &  	7030	  &  	45.04	  &  	8.95	  &  	2.33	   \\ 
1750+175	  &  	0	  &  	L	  &  	0.507	  &  	3700	  &  	45.45	  &  	8.60	  &  	2.34	   \\ 
\\															
{\it PG}	  &  		  &  		  &  		  &  		  &  		  &  		  &  		   \\ 
1226+023	  &  	3C 273	  &  	L	  &  	0.158	  &  	3520	  &  	46.10	  &  	8.88	  &  	2.51	   \\ 
1309+355	  &  	Ton 1565	  &  	L	  &  	0.182	  &  	2940	  &  	44.98	  &  	8.16	  &  	2.44	   \\ 
0947+396	  &  	K347-45	  &  	Q	  &  	0.206	  &  	4830	  &  	44.88	  &  	8.54	  &  	2.19	   \\ 
0953+414	  &  	K348-7	  &  	Q	  &  	0.234	  &  	3130	  &  	45.56	  &  	8.51	  &  	2.39	   \\ 
1001+054	  &  		  &  	Q	  &  	0.161	  &  	1740	  &  	44.87	  &  	7.65	  &  	2.46	   \\ 
1048+342	  &  		  &  	Q	  &  	0.167	  &  	3600	  &  	44.80	  &  	8.25	  &  	2.09	   \\ 
1114+445	  &  		  &  	Q	  &  	0.144	  &  	4570	  &  	44.73	  &  	8.42	  &  	2.36	   \\ 
1116+215	  &  	TON 1388	  &  	Q	  &  	0.176	  &  	2920	  &  	45.54	  &  	8.44	  &  	2.58	   \\ 
1202+281	  &  	GQ Com	  &  	Q	  &  	0.166	  &  	5050	  &  	44.66	  &  	8.47	  &  	2.24	   \\ 
1216+069	  &  		  &  	Q	  &  	0.332	  &  	5190	  &  	45.65	  &  	8.99	  &  	2.16	   \\ 
1322+659	  &  		  &  	Q	  &  	0.168	  &  	2790	  &  	44.92	  &  	8.09	  &  	2.03	   \\ 
1352+183	  &  	PB 4142	  &  	Q	  &  	0.151	  &  	3600	  &  	44.91	  &  	8.30	  &  	2.39	   \\ 
1411+442	  &  	PB1732	  &  	Q	  &  	0.090	  &  	2670	  &  	44.56	  &  	7.87	  &  	2.24	   \\ 
1427+480	  &  		  &  	Q	  &  	0.220	  &  	2540	  &  	44.89	  &  	7.99	  &  	2.31	   \\ 
1440+356	  &  	Mrk 478	  &  	Q	  &  	0.077	  &  	1450	  &  	44.53	  &  	7.33	  &  	2.30	   \\ 
1626+554	  &  		  &  	Q	  &  	0.132	  &  	4490	  &  	44.66	  &  	8.37	  &  	2.47	   \\ 
\\															
{\it G99}	  &  		  &  		  &  		  &  		  &  		  &  		  &  		   \\ 
RX J0022-34	  &  		  &  	Q	  &  	0.219	  &  	4110	  &  	44.97	  &  	8.45	  &  	2.18	   \\ 
ESO 242-G8	  &  		  &  	Q	  &  	0.059	  &  	3670	  &  	43.77	  &  	7.75	  &  	2.12	   \\ 
RX J0100-51	  &  		  &  	Q	  &  	0.062	  &  	3450	  &  	44.10	  &  	7.86	  &  	2.38	   \\ 
RX J0152-23	  &  		  &  	Q	  &  	0.113	  &  	3510	  &  	44.56	  &  	8.10	  &  	2.46	   \\ 
RX J0204-51	  &  		  &  	Q	  &  	0.151	  &  	5990	  &  	44.42	  &  	8.50	  &  	2.12	   \\ 
RX J0319-26	  &  		  &  	Q	  &  	0.079	  &  	4170	  &  	44.11	  &  	8.03	  &  	2.31	   \\ 
RX J0323-49	  &  		  &  	Q	  &  	0.071	  &  	2075	  &  	43.78	  &  	7.26	  &  	1.99	   \\ 
ESO 301-G13	  &  		  &  	Q	  &  	0.064	  &  	3180	  &  	44.08	  &  	7.78	  &  	2.37	   \\ 
VCV 0331-37	  &  		  &  	Q	  &  	0.064	  &  	2165	  &  	43.76	  &  	7.29	  &  	1.86	   \\ 
Fairall 1116	  &  		  &  	Q	  &  	0.059	  &  	4560	  &  	44.13	  &  	8.12	  &  	2.13	   \\ 
RX J0425-57	  &  		  &  	Q	  &  	0.104	  &  	2900	  &  	45.08	  &  	8.20	  &  	2.28	   \\ 
Fairall 303	  &  		  &  	Q	  &  	0.040	  &  	1720	  &  	43.39	  &  	6.90	  &  	1.78	   \\ 
RX J0435-46	  &  		  &  	Q	  &  	0.070	  &  	3820	  &  	43.52	  &  	7.66	  &  	2.04	   \\ 
RX J0437-47	  &  		  &  	Q	  &  	0.052	  &  	4215	  &  	43.98	  &  	7.98	  &  	2.01	   \\ 
RX J0438-61	  &  		  &  	Q	  &  	0.069	  &  	2410	  &  	44.07	  &  	7.54	  &  	1.91	   \\ 
CBS 126	  &  		  &  	Q	  &  	0.079	  &  	2850	  &  	44.31	  &  	7.80	  &  	2.20	   \\ 
Mkn 141	  &  		  &  	Q	  &  	0.042	  &  	4175	  &  	43.87	  &  	7.91	  &  	2.23	   \\ 
Mkn 734	  &  		  &  	Q	  &  	0.033	  &  	2230	  &  	43.94	  &  	7.40	  &  	2.28	   \\ 
IRAS 1239+33	  &  		  &  	Q	  &  	0.044	  &  	1900	  &  	43.91	  &  	7.25	  &  	2.31	   \\ 
RX J1646+39	  &  		  &  	Q	  &  	0.100	  &  	2160	  &  	43.85	  &  	7.33	  &  	1.99	   \\ 
RX J2232-41	  &  		  &  	Q	  &  	0.075	  &  	4490	  &  	43.67	  &  	7.87	  &  	2.29	   \\ 
RX J2245-46	  &  		  &  	Q	  &  	0.201	  &  	2760	  &  	45.41	  &  	8.32	  &  	2.46	   \\ 
RX J2248-51	  &  		  &  	Q	  &  	0.102	  &  	3460	  &  	44.50	  &  	8.07	  &  	1.99	   \\ 
MS 2254-37	  &  		  &  	Q	  &  	0.039	  &  	1545	  &  	43.84	  &  	7.04	  &  	2.41	   \\ 
RX J2258-26	  &  		  &  	Q	  &  	0.076	  &  	2815	  &  	44.00	  &  	7.63	  &  	2.08	   \\ 
RX J2349-31	  &  		  &  	Q	  &  	0.135	  &  	5210	  &  	44.32	  &  	8.33	  &  	2.31	   \\ 
\enddata

\tablecomments{
Black hole mass and [O~III] line width for AGN as described in the text.
Columns give 
(1)~reference (italicized) or object name, 
(2)~other name, 
(3)~radio loud (L) or quiet (Q),
(4)~redshift, 
(5)~FWHM of the broad H$\beta$ line,
(6)~continuum luminosity $(\nu L_\nu)$ at 5100~\AA\ rest wavelength,
(7)~black hole mass from equation (2), and
(8)~$\sigthree\ \equiv \wthree/2.35$.
References are D02: Dietrich et al. (2002), 
M99: McIntosh et al. (1999),
B96a: Brotherton (1996a), B96b: Brotherton (1996b), PG: this paper,
G99: Grupe et al. (1999).  See text for
discussion of continuum measurements.
}

\end{deluxetable}

\clearpage
\vskip 5pt \psfig{file=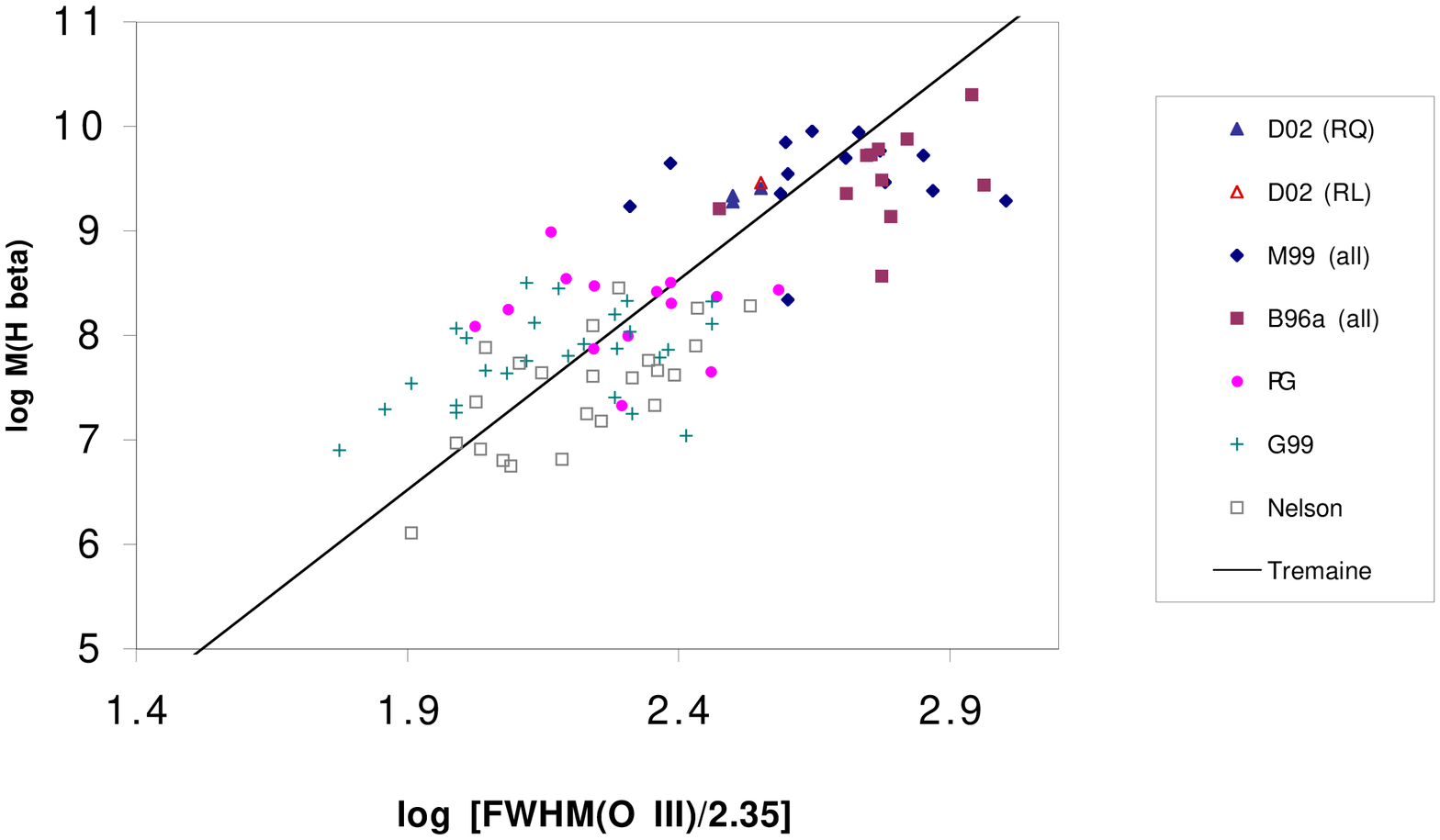,width=18cm,angle=0} 
\figcaption[fig1.ps]{
Black hole mass derived from \hbeta\ line width and continuum
luminosity versus width of the [O~III] line for radio quiet AGN (see
text).  Line is the \mbhsigstar\ relation from Tremaine \etal\ (2002),
given by equation (1); it is not a fit to the QSO data.  Shown here
are the ``full'' data sets from M99 and B96a, which include some
highly uncertain values (see text).  Key to data sources is given in
the footnote to Table 2: triangles--D02; diamonds--M99; squares--B96a;
dots--PG; crosses--G99.  Because of the scarcity of high redshift
points, we include as an open triangle the D02 object Q0256-0000,
which barely exceeds the threshold to be classified radio loud.
\label{fig1}}

\clearpage
\vskip 5pt \psfig{file=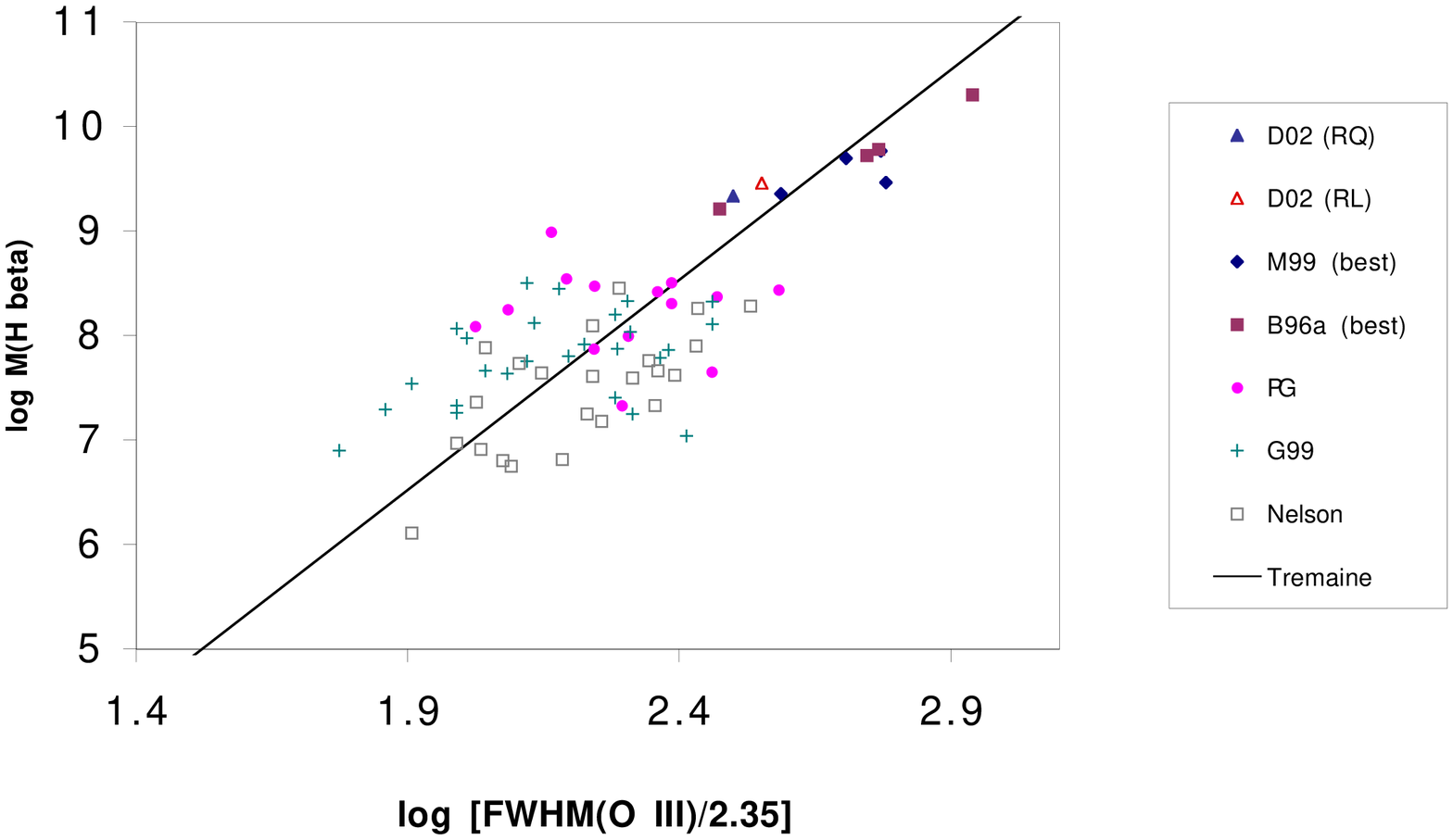,width=18cm,angle=0}
\figcaption[fig2.ps]{Same as Figure 2 but for the ``select'' data
sets from M99 and B96a.  See text for discussion of errors.
\label{fig2}}

\clearpage
\vskip 5pt \psfig{file=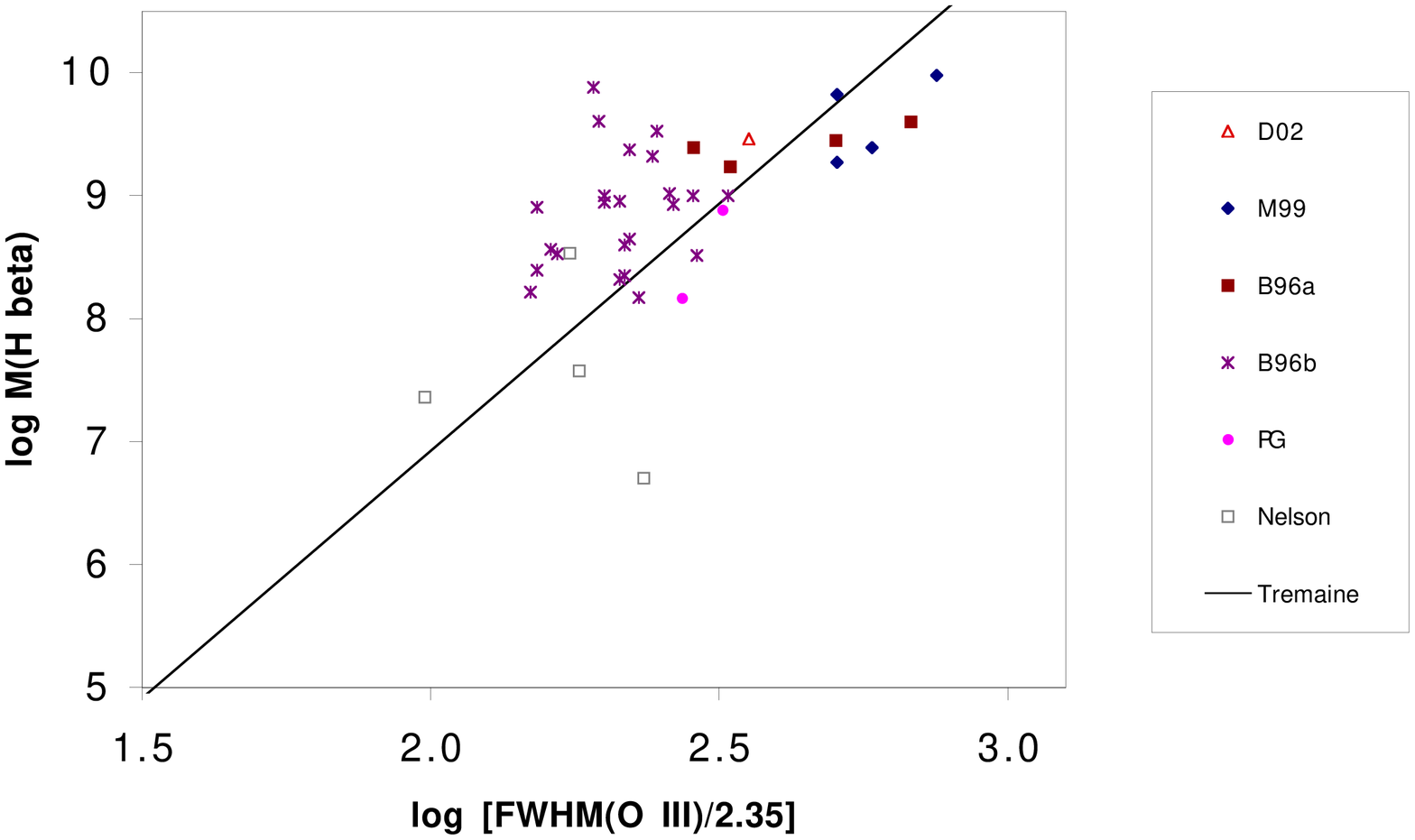,width=18cm,angle=0}
\figcaption[fig3.ps]{Same as Figure 2 but for radio loud AGN. B96b
data are shown as asterisks.
\label{fig3}}

\clearpage
\vskip 5pt \psfig{file=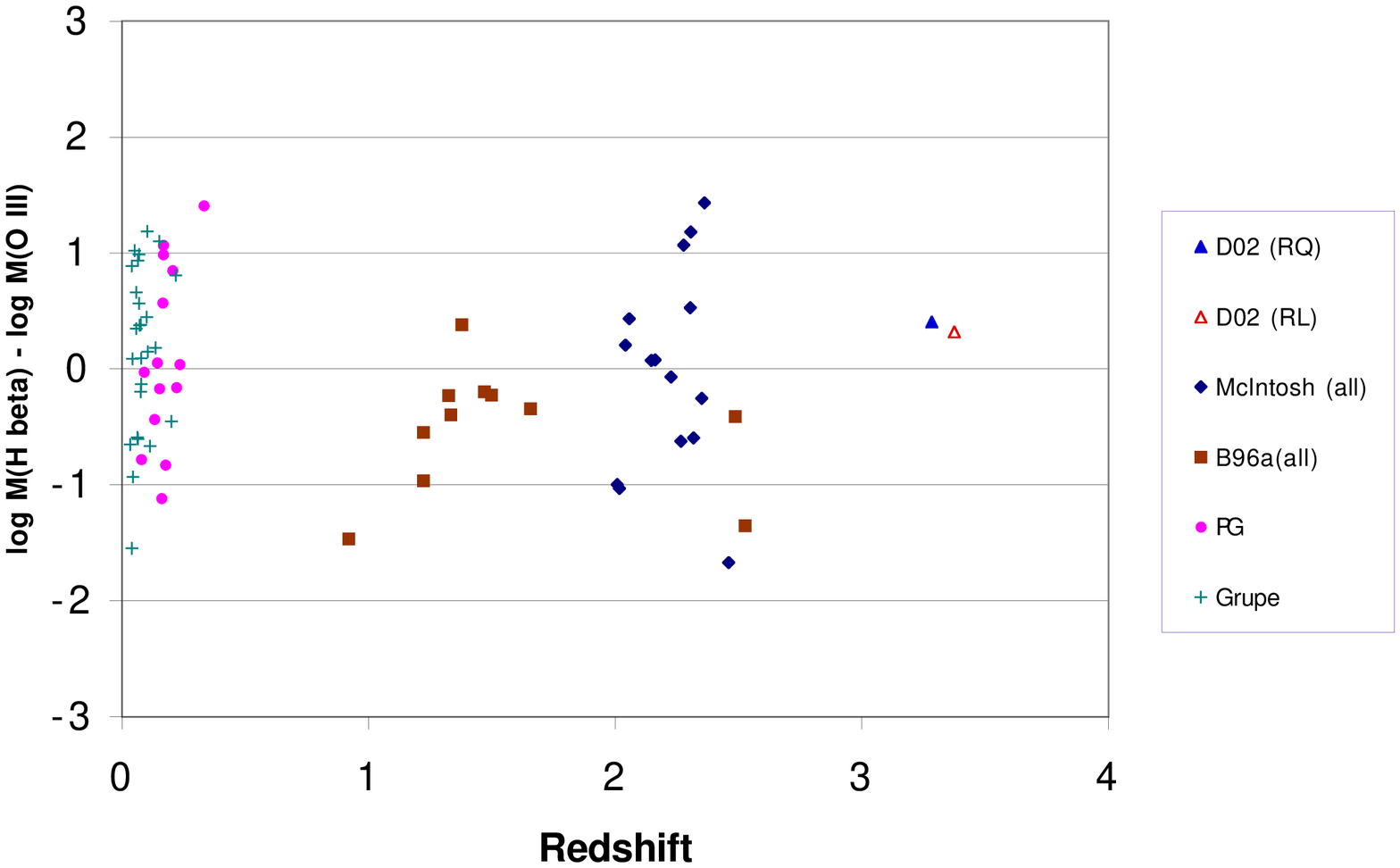,width=18cm,angle=0}
\figcaption[fig4.ps]{ Redshift dependence of the departure of measured
black hole mass from value expected from the \mbhsigstar\ relationship
(equation 1) and the measured [O~III] line width for radio quiet AGN
(see text).  Ordinate is $\Delta\, \log~M \equiv \log~M_{BH} -
\log~\mthree$, where $M_{BH}$ is the black hole mass derived from the
\hbeta\ line width and continuum luminosity (equation 2) and $\mthree$
is the black hole mass expected from the measured [O~III] line width
on the basis of equation (1).  Shown here are the ``full'' data sets
from M99 and B96a.  The high redshift QSOs do not differ significantly
in the mean from the low redshift objects in the relationship between
black hole mass and [O~III] line width.  This indicates little or no
change in the \mbhsigstar\ relation since the universe was 2 to 3
billion years old.
\label{fig4}}

\clearpage
\vskip 5pt \psfig{file=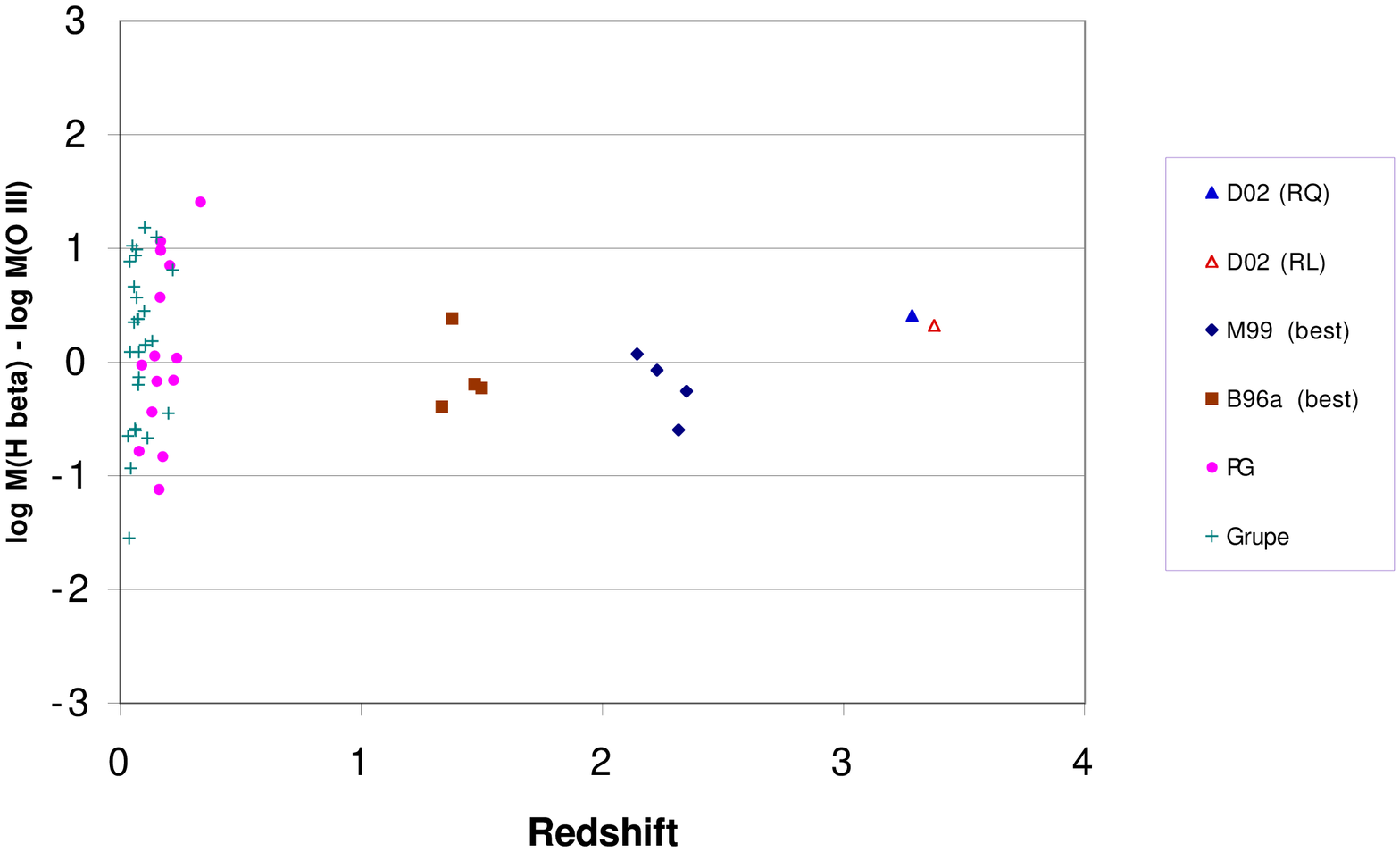,width=18cm,angle=0}
\figcaption[fig5.ps]{Same as Figure 4 but for the ``select'' data sets
from M99 and B96a.
\label{fig5}}

\clearpage
\vskip 5pt \psfig{file=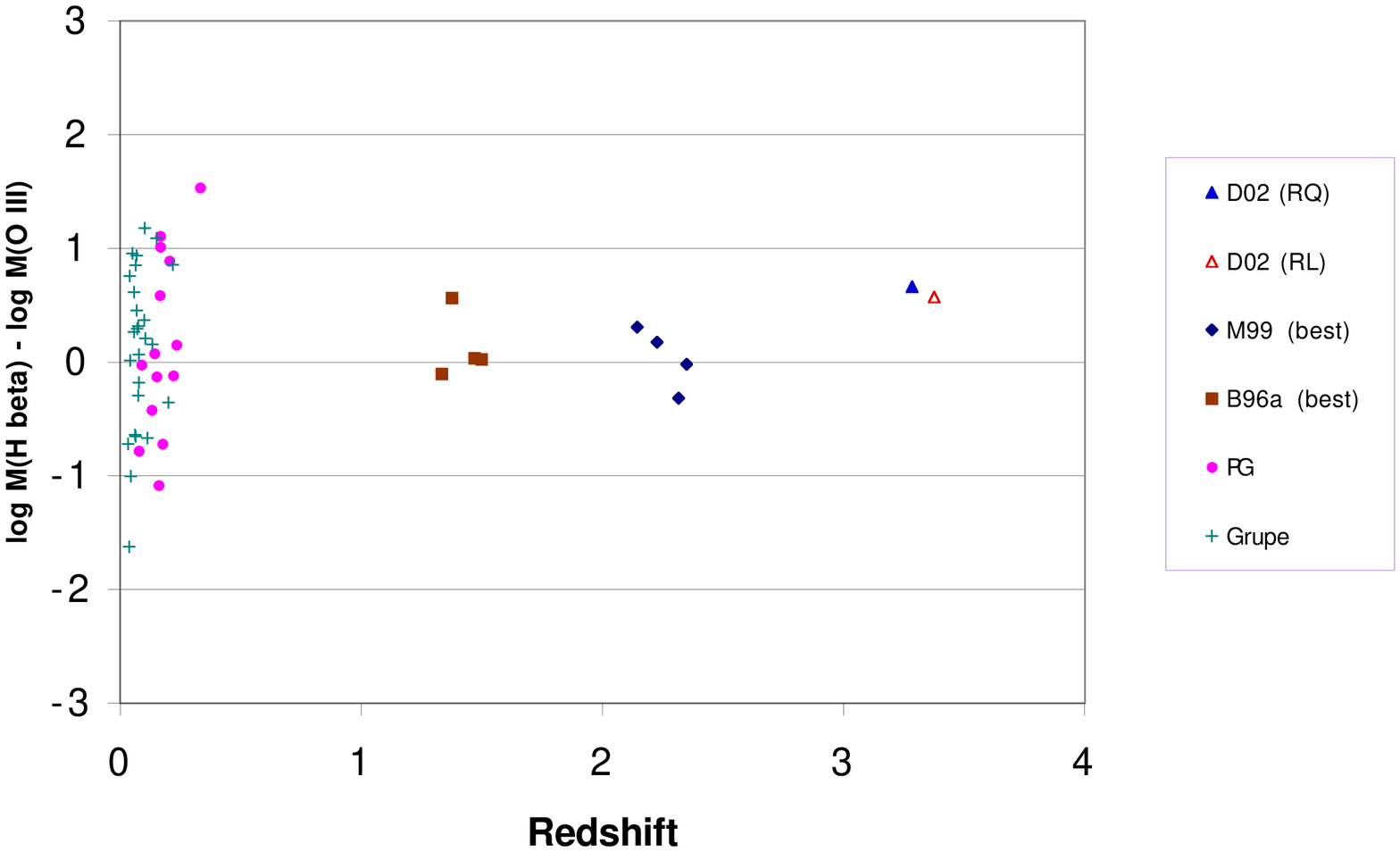,width=18cm,angle=0}
\figcaption[fig6.ps]{Same as Figure 5 but for $M_{BH} =
(10^{7.63}~\msun)v_{3000}^2 L_{44}^{0.61}$ (see text).
\label{fig6}}

\end{document}